\documentclass[a4paper,11pt,leqno]{amsart}
\usepackage{a4wide,color,array,graphicx}
\usepackage{mathtools}
\usepackage{hyperref}
\usepackage{cite}
\hypersetup{linktocpage,colorlinks, citecolor={magenta}}
\usepackage{amsmath,amssymb,amsthm}
\usepackage[english]{babel}
\usepackage[utf8]{inputenc}
\usepackage[cyr]{aeguill}
\usepackage{esint}
\usepackage{xargs}
\usepackage{lmodern}
\usepackage{bm}

\theoremstyle{plain}

\newtheorem{theo}{Theorem}
\newtheorem{prop}{Proposition}[section]

\newtheorem{coro}[prop]{Corollary}

\theoremstyle{definition}

\newtheorem{defi}[prop]{Definition}
\newtheorem{example}[prop]{Example}
\newtheorem{conjecture}[prop]{Conjecture}

\numberwithin{equation}{section}

\newcommand{\R}{\mathbb{R}}
\def\t0{\rightarrow 0} % Vers zéro
\def\ti{\rightarrow \infini} % Vers l'infini

\newcommand{\infini}{\infty}
\newcommand{\ep}{\varepsilon}

\newcommand{\hal}{\frac{1}{2}}

\def\div{\mathrm{div} \, } % Divergence
\def\1{\mathbf{1}} % Fonction caractéristique

\def \mc{\mathcal}
\def \ep{\epsilon}
\renewcommand{\epsilon}{\varepsilon}

\def\meseq{\mu_{V}} % Mesure d'équilibre
\def \ZNbeta{Z_{N,\beta}} % Fonction de partition

\def\({\left(}
\def\){\right)}

\def \W{\mathbb{W}} % Énergie d'une configuration de points
\def \bttW{\overline{\W}} % Énergie d'un processus ponctuel marqué
\def\config{\mathrm{Config}} % Espace des configurations de points

\def\P{\mathbb{P}} % "Vraies" mesures
\def \PNbeta{\P_{N, \beta}} % Mesure de Gibbs à \beta
\def \PgN2{\mathbf{P}_{N,2}} % Processus Gibbsien à N points et à \beta = 2..
\def \HN{\mathcal{H}_N}

\def \bERS{\overline{\mathsf{ent}}} % Entropie relative spécifique (sans marques)

\def \Poisson{{\Pi}}
\def \fbarbeta{\overline{\mathcal{F}}_{\beta}} % Rate function avec marque
\def\muv{\meseq}
\def\mueq{\meseq}
\def\I{\mathcal I}

\def\nab{\nabla}
\def\np{\nab^\perp}
\def\indic{\mathbf{1}}

\def\XXint#1#2#3{{\setbox0=\hbox{$#1{#2#3}{\int}$}
     \vcenter{\hbox{$#2#3$}}\kern-.5\wd0}}

\def\P{\mathbb{P}} % "Vraies" mesures
\def \PNbeta{\P_{N, \beta}} % Mesure de Gibbs à \beta
\def \PgN2{\mathbf{P}_{N,2}} % Processus Gibbsien à N points et à \beta = 2..
\def\g{\mathsf{g}}

\def \I{\mathcal{I}}
\def \LogU{\textsf{Log1}}
\def \LogD{\textsf{Log2}}
\def\Coul{\textsf{Coul}}

\def \C{\mathcal{C}}

\def\nab{\nabla}

\def\pa{{\partial}}

\def\ro{\rho}

\def\hal{\frac{1}{2}}

\def\I{{\mathcal{I}}}
\makeatletter
\def\namedlabel#1#2{\begingroup
    #2%
    \def\@currentlabel{#2}%
    \phantomsection\label{#1}\endgroup
}
\makeatother
\def \d{\mathsf{d}}
\def \s{\mathsf{s}}

\def\cd{\mathsf{c}_{\d}}

\def \bm{\begin{displaymath}}
\def \em{\end{displaymath}}
\def \be{\begin{equation}}
\def \ee{\end{equation}}
\def \beq*{\begin{equation*}}
\def \eeq*{\end{equation*}}
\def \ba{\begin{eqnarray}}
\def \ea{\end{eqnarray}}
\def \ba*{\begin{eqnarray*}}
\def \ea*{\end{eqnarray*}}

\long\def\replace#1{#1}

\replace{%

}

\begin{document}

\title{Systems of Points with Coulomb Interactions}

%%%%%%%%%%%%%%%%%%%%%%%%%%%%%%%%%%%%%%%%%%%%%%%%%%%%%%%%%%%%%%%%%%%%%
%    
%    Author information--add further authors as needed
%    
%%%%%%%%%%%%%%%%%%%%%%%%%%%%%%%%%%%%%%%%%%%%%%%%%%%%%%%%%%%%%%%%%%%%%
\author{Sylvia Serfaty}
\address{Courant Institute of Mathematical Sciences, New York University, 251 Mercer st, New York, NY 10012}
\email{serfaty@cims.nyu.edu}
%\date{November 22, 2017}

\maketitle
%%%%%%%%%%%%%%%%%%%%%%%%%%%%%%%%%%%%%%%%%%%%%%%%%%%%%%%%%%%%%%%%%%%%%
%    
%    Classification and abstract
%    
%%%%%%%%%%%%%%%%%%%%%%%%%%%%%%%%%%%%%%%%%%%%%%%%%%%%%%%%%%%%%%%%%%%%%

\begin{abstract}
Large ensembles of points with Coulomb interactions arise in various settings of condensed matter physics, classical and quantum mechanics, statistical mechanics, random matrices and even approximation theory, and give rise to a variety of questions pertaining to calculus of variations, Partial Differential Equations and probability. 
We will review these as well as ``the mean-field limit" results that allow to derive effective models and equations describing the system at the macroscopic scale. We then explain how to analyze the next order beyond the mean-field limit, giving information on the system at the microscopic level.  In the setting of statistical mechanics, this allows for instance to observe the effect of the temperature and to connect with crystallization questions.
\end{abstract}  
{\bf keywords: }Coulomb gases, log gases, mean field limits, jellium, large deviations, point processes, crystallization, Gaussian Free Field

\section{General setups }
We are interested in large systems of points with Coulomb-type interactions,  described through an energy of the form 
 \begin{equation}
 \label{nrj1} \HN (x_1, \dots, x_N)=\hal  \sum_{i\neq j} \g(x_i-x_j) + N \sum_{i=1}^N V(x_i).
 \end{equation}
 Here the points $x_i$ belong to the Euclidean space $\R^\d$, although it is also interesting to consider points on manifolds. 
 The interaction kernel $\g(x)$ is taken to be 
 \begin{align} 
\label{wlog2d} & (\LogD \ \text{case}) \quad \g(x) = - \log |x| , \quad  \text{in dimension } \d=2, \\
\label{kernel} & (\Coul \ \text{case}) \quad \g(x) = \frac{1}{|x|^{\d-2}}, \quad \text{ in dimension $\d \geq 3$}.
\end{align}
This is (up to a multiplicative constant) the Coulomb kernel in dimension $\d\ge 2$, i.e. the fundamental solution to the Laplace operator, solving 
\begin{equation}\label{eqg}
-\Delta \g = \cd \delta_0\end{equation}
where $\delta_0$ is the Dirac mass at the origin and $\cd $ is an explicit constant depending only on the dimension.
It is also interesting to broaden the study to the one-dimensional logarithmic case
\begin{equation}\label{wlog}  (\LogU \ \text{case}) \quad \g(x) = -\log |x| , \quad \text{in dimension } \d=1, \end{equation}
which is not Coulombian, and to more general Riesz interaction kernels of the form 
\begin{equation}\label{rieszgene}
\g(x)=\frac{1}{|x|^\s} \quad \s >0.\end{equation}
The one-dimensional Coulomb interaction with kernel $-|x|$ is also of interest, but we will not consider it as it has been extensively studied and understood, see  \cite{len1,len,kunz}.

Finally, we have included a possible external field or confining potential $V$, which is assumed to be regular enough and tending to $\infty$ fast  enough at $\infty$. The  factor $N$  in front of $V$ makes the total confinement energy  of the same order as the total  repulsion energy, effectively balancing them and  confining the system to a subset of $\R^\d$ of fixed size.  Other choices of scaling would lead to systems of very large or very small size as \medskip  $N \to \infty$.

The Coulomb interaction and the Laplace operator are obviously extremely important and ubiquitous in physics as the fundamental interactions of nature (gravitational and electromagnetic) are Coulombic.  
 Coulomb was a French engineer and physicist working in the late 18th century,  who did a lot of work on applied mechanics (such as  modeling friction and torsion) and is most famous for his theory of electrostatics and magnetism. He is the first one who postulated that  the force exerted by charged particles is proportional to the inverse distance squared, which corresponds in dimension $\d= 3$ to the gradient of the Coulomb potential energy $\g(x)$ as above.  More precisely he wrote in  \cite{cou}
`` It follows therefore from these three tests, that the repulsive force that the two balls  [which were] electrified with the same kind of electricity  exert on each other, follows the inverse proportion of the square of the distance."
 He developed a method based on systematic use of  mathematical calculus (with the help of suitable approximations)  and mathematical modeling (in contemporary terms)  to  predict physical behavior, systematically  comparing the results with the measurements of the experiments he was designing and conducting himself.  As such, he is considered as a pioneer of the ``mathematization" of physics and in trusting fully the capacities of mathematics to transcribe physical \medskip phenomena \cite{bw}.

Here we are more specifically focusing on  Coulomb interactions between points, or in physics terms, discrete point charges.
There are several mathematical problems that are interesting to  study, all in the asymptotics of $N \to \infty$ :
\begin{enumerate}
\item[\textbf{(1)}] understand minimizers and possibly critical points of \eqref{nrj1} ;
\item[\textbf{(2)}] understand the statistical mechanics of systems with energy $\HN$ and inverse temperature $\beta>0$, governed by the so-called Gibbs measure 
\begin{equation}\label{gibbs}
d\PNbeta(x_1, \dots, x_N)= \frac{1}{\ZNbeta} e^{-\beta \HN(x_1, \dots, x_N) } dx_1\dots dx_N.\end{equation}
Here, as postulated by statistical mechanics, $\PNbeta$ is the density of  probability of observing the system in the configuration $(x_1, \dots, x_N)$ if the number of particles is fixed to $N$ and  the inverse of the temperature is $\beta>0$. 
The constant $\ZNbeta$ is called the ``partition function" in physics, it is the normalization constant that makes $\PNbeta$ a probability measure, \footnote{One does not know how to explicitly compute the integrals \eqref{defZ} except in the particular case of \eqref{wlog}  for specific $V$'s where they are called Selberg integrals (cf. \cite{mehta,forrester})} i.e. 
\begin{equation}
\label{defZ}
\ZNbeta= \int_{(\R^\d)^N} e^{-\beta \HN(x_1, \dots, x_N) } dx_1\dots dx_N,\end{equation} 
where  the inverse temperature $\beta=\beta_N$ can be taken to depend on $N$, as there are several interesting scalings of $\beta$ relative to $N$;

\item[\textbf{(3)}] understand dynamic evolutions associated to \eqref{nrj1}, such as 
the gradient flow of $\HN$ given by the system of coupled ODEs 
\begin{equation}
\label{gf}\dot{ x_i} = - \frac{1}{N} \nab_{i} \HN(x_1, \dots,x_N),
\end{equation}
 conservative  dynamics given  by the systems of ODEs
\begin{equation}
\label{hf}
\dot{ x_i} =\frac{1}{N} \mathbb{J} \nab_i \HN(x_1, \dots, x_N)\end{equation} where $\mathbb{J}$ is an antisymmetric matrix (for example a rotation by $\pi/2$ in dimension 2), 
or the Hamiltonian dynamics given by Newton's law
\be \label{newton}
\ddot{ x_i}= - \frac{1}{N} \nab_i \HN(x_1, \dots, x_N);\ee

\item[\textbf{(4)}] understand the previous dynamic evolutions   with  temperature $\beta^{-1}$ in the form of an added noise (Langevin-type equations) such as 
\begin{equation}\label{noise1}
d x_i = -\frac{1}{N} \nab_i \HN(x_1, \dots, x_N) dt + \sqrt{\beta^{-1}} dW_i\end{equation} with $W_i$ independent Brownian motions, or 
\be\label{noise2}
d x_i = \frac{1}{N}\mathbb{J} \nab_i \HN(x_1, \dots, x_N) dt + \sqrt{\beta^{-1}} dW_i \ee
with $\mathbb{J}$ as above, or 
\be \label{noise3}
d x_i = v_i dt \qquad dv_i=  -\frac{1}{N} \nab_i \HN(x_1, \dots, x_N) dt + \sqrt{\beta^{-1}} dW_i.\ee

\end{enumerate}

From a mathematical point of view, the study of such systems touches on the fields of analysis (Partial Differential Equations and calculus of variations, approximation theory) particularly for \textbf{(1)}-\textbf{(3)}-\textbf{(4)}, probability (particularly for \textbf{(2)}-\textbf{(4)}), mathematical physics, and even geometry  (when one considers such systems on manifolds or with curved geometries). Some of the crystallization questions they lead to also overlap with number theory as we will see below.

In the sequel we will mostly focus on the stationary settings \textbf{(1)} and \textbf{(2)}, while mentioning more briefly some results about \textbf{(3)} and \textbf{(4)}, for which many questions remain open. Of course these various points are not unrelated, as for instance
the Gibbs measure \eqref{gibbs} can also be seen as an invariant measure for dynamics of the form \eqref{newton} \medskip  or \eqref{noise1}.

The plan of the discussion is as follows:  in the next section we review various motivations for studying such questions, whether from physics or within mathematics.  In Section \ref{sec2} we turn to the so-called ``mean-field" or leading order description of systems \textbf{(1)} to \textbf{(4)}  and review the standard questions and  known results. We emphasize that this part can be extended to general interaction kernels $\g$, starting with regular (smooth) interactions which are in fact the easiest to treat.  
In Section \ref{sec3}, we discuss questions that can be asked and results that can be obtained at the next order level of expansion of the energy. This has only been tackled for problems \textbf{(1)} and \textbf{(2)}, and the specificity of the Coulomb interaction becomes important then.

\section{Motivations}\label{sec1}

It is in fact impossible to list all possible topics in which such systems arise, as they are really numerous. We will attempt to give a short, necessarily biased, list of examples, with possible pointers to the relevant literature.

\subsection{Vortices in condensed matter physics and fluids}

\begin{figure}[ht!]
\begin{center}
\includegraphics[width=5cm]{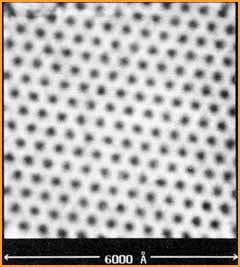}
\caption{Abrikosov lattice, H. F. Hess et al. Bell Labs
{\it Phys. Rev. Lett.} 62, 214 (1989)}
\label{fig32}

\end{center}

\end{figure}

In superconductors with applied magnetic fields, and in rotating  superfluids and Bose-Einstein condensates, one observes  the occurrence of quantized ``vortices" (which are local point defects of superconductivity or superfluidity, surrounded by a current loop). The vortices repel each other, while being confined together by the effect of the magnetic field or rotation,  and the result of the competition between  these two effects is that, as predicted by Abrikosov \cite{a}, they arrange themselves in a particular {\it triangular lattice} pattern, called {\it Abrikosov lattice}, cf. Fig. 
\ref{fig32} (for more pictures, see {\tt www.fys.uio.no/super/vortex/}).
Superconductors and superfluids are modelled by the celebrated Ginzburg-Landau energy \cite{gl}, which in simplified form \footnote{The complete form for superconductivity contains a gauge-field, but we omit it here for simplicity.} can be written 
\be \label{gl} 
\int |\nab \psi|^2 + \frac{(1-|\psi|^2)^2}{2\ep^2}\ee
where $\psi$ is a complex-valued unknown function (the ``order parameter" in physics) and $\ep $ is a small parameter,
and gives rise to the  associated Ginzburg-Landau equation
\be\label{gle}
\Delta \psi+ \frac{1}{\ep^2} \psi(1-|\psi|^2)=0\ee and its dynamical versions, the heat flow 
\be\label{glhf}
\pa_t \psi=\Delta \psi+ \frac{1}{\ep^2} \psi(1-|\psi|^2)\ee
and Schr\"odinger-type flow (also called the Gross-Pitaevskii equation)
\be\label{gls}
i\pa_t \psi=\Delta \psi+\frac{1}{\ep^2} \psi(1-|\psi|^2).\ee

When restricting to a two-dimensional situation, it can be shown  rigorously (this was pioneered by \cite{bbh} for \eqref{gl} and extended to the full gauged model \cite{bethriv,livre,ssgl}) that the minimization of \eqref{gl} can be reduced, in terms of the vortices and as $\ep \to 0$, to the minimization of an energy of the form \eqref{nrj1} in the case \eqref{wlog2d} (for a formal derivation, see also \cite[Chap. 1]{ln}) and  this naturally leads to the question of understanding the connection between minimizers  of \eqref{nrj1} + \eqref{wlog2d} and the Abrikosov triangular lattice.
Similarly, the dynamics of vortices  under \eqref{glhf} can be formally reduced to \eqref{gf}, respectively under \eqref{gls} to \eqref{hf}. This was established formally for instance in \cite{pr,E} and proven for a fixed number of vortices $N$ and  in the limit $\ep \to 0$ in \cite{li,js,cj1,cj2,lx,lx2,bjs}  until the first collision time and in \cite{bos1,bos2,bos3,s2} including after collision.

Vortices also arise in classical fluids, where in contrast with what happens in superconductors and superfluids, their charge is not quantized. In that context the energy \eqref{nrj1}+\eqref{wlog2d} is sometimes called the  Kirchhoff energy 
 and the system \eqref{hf} with $\mathbb{J}$ taken to be a rotation by $\pi/2$, known as the point-vortex system, corresponds to the dynamics of idealized vortices in an incompressible fluid whose statistical mechanics analysis was initiated by Onsager, cf. \cite{es} (one of the motivations for studying \eqref{noise2} is precisely to understand fluid turbulence as he conceived).
  It has thus been quite  studied as such, see \cite{mp} for further reference. 
  The study of evolutions like \eqref{newton} is also motivated by plasma physics in which the interaction between ions is Coulombic, cf. \cite{jabin}.

%Finally, systems governed by \eqref{nrj1} can be seen as classical instead of quantum  toy model for matter: for instance, Gamov's ``liquid drop model" for the atomic nucleus of Gamov (see review) is also a simplified model for electrons and atoms, and in some regime where one phase is in large majority, it can be reduced to a system of points interacting like \ref{nrj1} (see for instance \cite{gms} and references therein).

\subsection{Fekete points and approximation theory}
Fekete points arise in interpolation theory as the points minimizing  interpolation errors for numerical integration \cite{safftotik}.   More precisely, if one is looking for  $N$ interpolation points $\{x_1, \dots, x_N\}$ in $K$ such that the  relation
\[
\int_K f(x) dx = \sum_{j=1}^N w_j f(x_j)
\]
is exact when $f$ is any  polynomial of degree $\le N-1$, one sees that one needs to compute the coefficients $w_j$ such that  $\int_{K} x^k = \sum_{j=1}^N w_j x_j^k$ for $0 \leq k \leq N-1$, and this computation is easy if one knows to invert the Vandermonde matrix of the  $\{x_j\}_{j=1 \dots N}$. The numerical stability of this operation is as large as the  \textit{condition number} of the  matrix, i.e. as the Vandermonde determinant of the $(x_1, \dots, x_N)$. The  points that minimize the maximal interpolation error for general functions are easily shown to  be the Fekete points, defined as those that maximize
$$\prod_{i\neq j} |x_i-x_j|$$
 or equivalently minimize
$$-\sum_{i\neq j} \log |x_i-x_j|.$$

They are often studied on manifolds, such as the $\d$-dimensional sphere.
In Euclidean space, one also considers ``weighted Fekete points" which maximize
$$\prod_{i< j} |x_i-x_j| e^{-N\sum_i V(x_i)}$$
or equivalently minimize
$$-\hal \sum_{i\neq j} \log |x_i-x_j| + N\sum_{i=1}^N V(x_i)$$
which in dimension $2$ corresponds exactly to the minimization of  $\HN$ in the particular case \LogD.  They also happen to be zeroes of orthogonal polynomials, see \cite{simon}.

Since $-\log |x|$ can be obtained as $\lim_{s\to 0} \frac{1}{s}(|x|^{-s}-1)$, there is also interest in studying ``Riesz $\s$-energies", i.e. the minimization of 
\begin{equation}\label{rieszs}
\sum_{i\neq j} \frac{1}{|x_i-x_j|^\s}\end{equation}
for all possible $\s$, hence a motivation for \eqref{rieszgene}. 
For these aspects, we refer to the  the review papers \cite{sk,bhs}  and references therein.

Varying $\s$ from $0$ to $\infty$ connects Fekete points to the optimal sphere  packing problem, which formally corresponds to the minimization of \eqref{rieszs} with $\s=\infty$. 

\begin{figure}
\includegraphics[height=3.5cm]{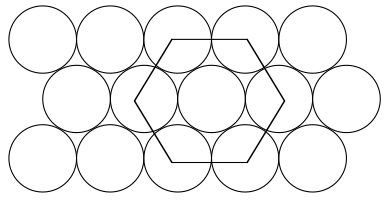}
\caption{The triangular lattice solves the sphere packing problem in dimension 2}\label{figsp}
\end{figure}

The  optimal sphere packing problem  has been solved in 1, 2 and 3 dimensions, as well as in dimensions 8 and 24 in the recent breakthrough  \cite{via,ckrmv}  (we refer the reader to the nice presentation in \cite{cohn} and the review \cite{sloane}). The solution in dimension 2 is the triangular lattice \cite{ft} (i.e. the same as the Abrikosov lattice, see Figure \ref{figsp}),
 in dimension 3 it  is the FCC (face-centered cubic) lattice \cite{hales},  in dimension 8 the $E_8$ lattice \cite{via}, and in dimension 24 the Leech lattice \cite{ckrmv}.

 In other dimensions, the solution is in general not known and it is expected that in high dimension, where the problem is important for error-correcting codes,  it  is {\it not} a lattice (in dimension 10 already, the  so-called ``Best lattice", a non-lattice competitor, is known to beat the lattices), see \cite{conwaysloane} for these aspects.

\subsection{Statistical mechanics and quantum mechanics}

The ensemble given by \eqref{gibbs} in the  \LogD \ case is called in physics a two-dimensional \textit{Coulomb gas} or \textit{one-component plasma} and is a classical ensemble of statistical mechanics (see e.g. \cite{aj,jlm,janco,sm,kiessling,kiesslingspohn}). 
The Coulomb case with $\d = 3$ can be seen as a toy (classical) model for matter (see e.g. \cite{PenroseSmith,jlm,LiLe1,LN}).  
 Several additional motivations come from quantum mechanics. Indeed,
 the Gibbs measure of the two-dimensional Coulomb gas 
  happens to be directly related to the Laughlin wave-function  in the fractional quantum Hall effect \cite{Gir,stormer} : this is the ``plasma analogy", cf. \cite{laughlin1,laughlin2,laughlin3}, and for recent mathematical progress using this correspondence, cf. \cite{RSY2,rougerieyngvason,lry}.   
For $\beta=2$ it also arises as the wave-function density of the ground state for 
the system of $N$ non-interacting fermions confined to a plane with a perpendicular
magnetic field \cite[Chap. 15]{forrester}.
The 1-dimensional log gas \LogU \ also arises as the wave-function density in several  exactly solvable quantum mechanics systems: examples are the Tonks-Girardeau   model of ``impenetrable"  Bosons 
\cite{gwt,ffgw},  the Calogero-Sutherland  quantum many-body Hamiltonian  \cite{fjm,forrester} and finally
the density of the many-body wave function of  non-interacting fermions in a harmonic trap  \cite{ddms}.  It also arises in several non-intersecting paths models from probability, cf. \cite{forrester}.

The general Riesz case \eqref{rieszgene} can be seen as a generalization of the Coulomb case, motivations for studying Riesz gases are numerous in the physics literature (in solid state physics, ferrofluids, elasticity), see for instance \cite{mazars,bbdr,CDR,To}, they can also correspond to systems with Coulomb interaction constrained to a lower-dimensional subspace : for instance in the quantum Hall effect, electrons confined to a two-dimensional plane interact via the three-dimension Coulomb kernel.

In all cases of interactions, the systems governed by the Gibbs measure $\PNbeta$ are considered as difficult systems of statistical mechanics because the interactions are truly long-range, singular, and the points are not constrained to live on a lattice.

As always in statistical mechanics \cite{huang}, one would like to understand if there are phase-transitions for particular values of the (inverse) temperature $\beta$ in the large volume limit. For the systems studied here, one may expect, after a suitable blow-up of the system, what physicists call a liquid for small $\beta$, and a crystal for large $\beta$. 
The meaning of crystal in this instance is not to be taken literally as a lattice, but rather as a system of points whose 2-point correlation function
$\ro^{(2)}(x,y)$ defined as the probability to have jointly one point at $x$ and one point at $y$ (see Section \ref{secmf})
  does not decay too fast as $x-y \to \infty$.
A phase-transition  at finite $\beta$ has been conjectured in the physics literature for the \LogD \ case (see e.g. \cite{bst,caillol1982monte,choquard1983cooperative}) but its precise nature is still unclear (see e.g. \cite{stishov} for a discussion).

\subsection{Two component plasma case} \label{2comp}
The two-dimensional ``one component plasma", consisting of positively charged particles, has a ``two-component" counterpart which consists in $N$ particles $x_1, \dots , x_N$  of charge $+1$ and $N$ particles $y_1, \dots , y_N$ of charge $-1$ interacting logarithmically, with energy 
$$\HN(x_1, \dots, x_N, y_1, \dots, y_N)= - \sum_{i\neq j} \log |x_i-x_j|- \sum_{i\neq j} \log |y_i-y_j|+  \sum_{i, j} \log |x_i-y_j|$$
and the Gibbs measure $$\frac{1}{\ZNbeta} e^{-\beta \HN( x_1, \dots, x_N, y_1, \dots, y_N ) } dx_1 \dots dx_N \, dy_1 \dots dy_N.$$
Although the energy is unbounded below (positive and negative points attract), the Gibbs measure is well defined for $\beta$ small enough, more precisely the partition function converges for $\beta<2$. The system is then seen to form dipoles of oppositely charged particles which attract but do not collapse, thanks to the thermal agitation. 
The two-component plasma is interesting due to its close relation to two important theoretical physics models: the XY model and the sine-Gordon model (cf. the review \cite{spencer}), which exhibit  a Berezinski-Kosterlitz-Thouless phase transition \cite{bietenholzgerber} consisting in the binding of these ``vortex-antivortex" dipoles.  For further reference, see \cite{Frohlich,DeutschLavaud,FS,GunPan}.

\subsection{Random matrix theory}
The study of \eqref{gibbs} has attracted a lot of attention due to its connection with random matrix theory (we refer to \cite{forrester} for a comprehensive treatment).  Random matrix theory (RMT) is a relatively old theory, pionereed by statisticians and physicists such  as Wishart,  Wigner and Dyson, and originally motivated by  the study of sample covariance matrices for the former and the understanding of the spectrum of heavy atoms for the two latter, see \cite{mehta}. For more recent mathematical reference see \cite{agz,deift,forrester}. The main question asked by RMT is~: what is the law of the spectrum of a large random matrix~? 
As first noticed in the foundational papers of  \cite{wigner,dyson}, in the particular cases \eqref{wlog} and \eqref{wlog2d}  the Gibbs measure \eqref{gibbs}  corresponds in some particular instances to the joint law of the eigenvalues (which can be computed algebraically) of some famous random matrix ensembles:
\begin{itemize}

\item for \LogD , $\beta=2$ and $V(x)=|x|^2$, \eqref{gibbs} is the law of the (complex)  eigenvalues of an $N\times N$ matrix where the entries are chosen to be normal Gaussian i.i.d.  This is called the  Ginibre ensemble  \cite{ginibre}.

\item for \LogU , $\beta=2$ and $V(x)= x^2/2$, \eqref{gibbs} is the law of the (real) eigenvalues of an $N\times N$ Hermitian matrix with complex normal Gaussian iid entries. This is called the Gaussian Unitary Ensemble.

\item for \LogU  , $\beta=1$ and $V(x)=x^2/2$, \eqref{gibbs} 
 is the law of the (real) eigenvalues of an $N\times N$ real symmetric  matrix with normal Gaussian iid entries. This is called the Gaussian Orthogonal Ensemble.

\item for \LogU ,  $\beta=4$ and $V(x)=x^2/2$, \eqref{gibbs} is the law of the eigenvalues of an $N\times N$  quaternionic symmetric  matrix with normal Gaussian iid entries. This is called the Gaussian Symplectic Ensemble. 
\item the general-$\beta$ case of \LogU \ can also be represented, in a slightly more complicated way, as a random matrix ensemble \cite{de,kn}.

\end{itemize}
 One thus observes in these ensembles the phenomenon of ``repulsion of eigenvalues": they repel each other logarithmically, i.e.  like two-dimensional Coulomb particles. 
 
 The stochastic evolution \eqref{noise1} in the case \LogU\  is (up to proper scaling) the Dyson Brownian motion, which is of particular importance in random matrices since the GUE process is the invariant measure for this evolution, it has served to prove universality for the statistics of eigenvalues of general Wigner matrices, i.e. those with iid but not necessarily Gaussian entries, see  \cite{ey} (and \cite{tv} for another approach).
   
 For the \LogU \ and \LogD \  cases, at the  specific temperature $\beta=2$, the law \eqref{gibbs} acquires a  special algebraic feature : it becomes a {\it determinantal} process, part of a wider class of processes (see \cite{bkpv,borodin}) for which the correlation functions are explicitly given by certain  determinants. This allows for many explicit  algebraic computations, and is part of {\it integrable probability} on which there is a large literature \cite{borgor}.   

   \subsection{Complex geometry and theoretical physics}
Coulomb systems and higher-dimensional analogues involving powers of determinantal densities are 
also of interest to geometers   as a way to construct K\"ahler-Einstein metrics with negative Ricci curvature on complex manifolds, cf. \cite{berman,bbn}.

   Another  important motivation is the construction of Laughlin states for the Fractional Quantum Hall effect on complex manifolds, which effectively reduces to the study of a two-dimensional Coulomb gas on a manifold. The coefficients in the expansion of the (logarithm of the) partition function have interpretations  as geometric invariants, cf. for instance \cite{klevtsov}. 
     
  \section{The mean field limits and macroscopic behavior}\label{sec2}
  \subsection{Questions}\label{secmf}
  The first question that naturally arises is to understand the limit as $N\to \infty$ of the {\it empirical measure} defined by 
\footnote{Note that the configurations contain $N$ points which also implicitly depend on   $N$ themselves, but we do not keep track of this dependence for the sake of lightness of notation.}
  \be \label{em}
  \mu_N :=\frac{1}{N} \sum_{i=1}^N \delta_{x_i}\ee
  for configurations of points that minimize the energy \eqref{nrj1},  critical points,  solutions of the evolution problems, or 
typical configurations under the Gibbs measure \eqref{gibbs}, thus hoping to derive effective equations or minimization problems that describe the average or mean-field behavior of the system. The term mean-field refers to the fact that, from the physics perspective, each particle feels the collective  field  generated by all the other particles, averaged by dividing it by the number of particles. That collective field is $\g* \mu_N$, except that it is singular at each particle, so to evaluate it at $x_i$ one first has to remove the contribution of $x_i$ itself.

Another point of view is that of correlation functions. One may denote by 
\be \rho_N^{(k)}(x_1, \dots, x_k)\ee
the $k$-point correlation function, which is the  probability density (for each specific problem) of observing  a particle at $x_1$,
a particle at $x_2$, $\dots$, and a particle at $x_k$ (these functions should of course be symmetric with respect to permutation of the labels). For instance, in the case \eqref{gibbs}, 
$\rho_N^{(N)}$ is simply $\PNbeta$ itself, and the $\rho_N^{(k)}$ are its marginals (obtained by integrating $\PNbeta$ with respect to all its variables but $k$).
One then wants to understand the limit as $N\to \infty$ of each $\ro_N^{(k)}$, with fixed $k$. Mean-field results will typically imply that  the limiting
$\rho^{(k)}$'s have a factorized form
\be \rho^{(k)} (x_1 , \dots, x_k)= \mu(x_1) \dots \mu(x_k)\ee
for the appropriate $\mu$ which is also equal to $\ro^{(1)}$.  This is called {\it molecular chaos} according to the terminology introduced by Boltzmann, and can be interpreted as the particles becoming independent in the limit. 
When looking at the dynamic evolutions of problems \textbf{(3)} and \textbf{(4)}, starting from initial data for which $\rho^{(k)}(0, \cdot)$ are in such a factorized form, one asks whether this remains true for $\rho^{(k)} (t, \cdot)$ for $t>0$,  if so this is called {\it propagation of (molecular) chaos}.
It turns out that the convergence of  the empirical measure \eqref{em} to a limit $\mu$ and the fact that each $\ro^{(k)}$ can be put in factorized form are essentially equivalent, see \cite{hm,golse} and references therein --- ideally, one would also like to find quantitative rates of convergences in $N$, and they will typically deteriorate as $k$ gets large.  
In the following we will focus on the mean-field convergence approach, via the empirical measure. 

In  the case of minimizers \textbf{(1)}, a major question is to obtain an expansion as $N \to \infty$ for $\min \HN$. In the setting of manifolds, the coefficients in such an expansion have geometric interpretations. In the same way,
in the statistical mechanics setting \textbf{(2)}, one searches for expansions as $N \to \infty$ of the so-called {\it free energy} $-\beta^{-1} \log \ZNbeta$. The free energy encodes 
 a lot of the physical quantities of the system. For instance, points of non-differentiability of $\log \ZNbeta$ as a function of $\beta$ are interpreted as phase-transitions. 
 
We will see below that understanding  the mean-field behavior of the system essentially amounts to understanding the leading order term in the large $N$ expansion of the minimal energy or respectively the free energy, while understanding the next order term in the expansion essentially amounts to understanding the next order (or fluctuations) of the system.

 \subsection{The equilibrium measure}
The leading order behavior of $\HN$ is related to  the functional
\begin{equation} \label{definitionI}
\I_V(\mu) := \hal\iint_{\R^\d \times \R^\d} \g(x-y) d\mu(x)d\mu(y) + \int_{\R^\d} V(x) d\mu(x)
\end{equation}
defined over the space $\mc{P}(\R^\d)$ of probability measures on $\R^\d$ (which may also take the value $+ \infty$).
This is something one may naturally expect since $\I_V(\mu)$ appears as the continuum  version of the discrete energy  $\HN$. From the point of view of statistical mechanics, $\I_V$ is the mean-field limit energy of  $\HN$, while from the point of view of probability, $\I_V$ plays the role of a {\it rate function}.

 Assuming some lower semi-continuity of $V$ and that it grows faster than $\g$ at $\infty$, it was shown in 
 \cite{frostman} that the minimum of $\I_V$ over $\mc{P}(\R^\d)$ exists, 
 is finite and is achieved by a unique $\meseq$ (unique by strict convexity of $\I_V$), which  has  compact support and a density, and is  uniquely characterized by the fact that there exists a constant $c$ such that 
\begin{equation}
\label{EulerLagrange}
\left\lbrace
\begin{array}{cc} h^{\meseq} + V \geq c & \mbox{ in } \R^\d  \vspace{3mm} \\ 
 h^{\meseq} + V= c & \mbox{ in the support of }\meseq \end{array} \right.
\end{equation}
where \be \label{defhmu0}
h^{\meseq}(x) := \int_{\R^\d} \g(x - y) d\meseq(y)= \g* \meseq
\ee is the ``electrostatic" potential generated by $\meseq$.

%, which in the Coulomb case solves 
%\begin{displaymath}
%- \Delta h^{\meseq} =\cd \meseq
%\end{displaymath} 
This measure $\meseq$ is  called the (Frostman) \textit{equilibrium measure}, and the result is true for more general repulsive  kernels than Coulomb, for instance for all regular kernels or inverse powers of the distance which are integrable.

\begin{example} When $\g$ is the Coulomb kernel, applying the Laplacian on both sides of \eqref{EulerLagrange} gives that, in the interior of the support of the equilibrium measure, if $V\in C^2$, \be \label{densmu0}
\cd \meseq = \Delta V
\ee
 i.e. the density of the measure on the interior of its support is given by $\frac{\Delta V}{\cd}$.  For example if $V$ is quadratic, this density is constant  on the interior of its support. If $V(x)=|x|^2 $ then by 
symmetry $\meseq$ is the indicator function of a ball (up to a multiplicative factor), this is known as the {\it circle  law} for the Ginibre ensemble in the context of Random Matrix Theory.  An illustration of the convergence to this circle law can be found in Figure \ref{fig11}.
In dimension $\d=1$, with $\g =-\log |\cdot |$ and $V(x) = x^2$, the equilibrium measure is $\meseq (x)= \frac{1}{2\pi} \sqrt{4-x^2} \mathbf{1}_{|x|\leq 2}$, which corresponds in the context of RMT (GUE and GOE ensembles) to the famous {\it   Wigner semi-circle law}, cf. \cite{wigner,mehta}.
\end{example}
%We will always assume that $\meseq$ is a measure with a H\"older continuous density on its support, we abuse notation by denoting its density $\meseq(x)$ and we also assume that its support $\Supp$ is a compact set with a nice boundary. We allow for several connected components of $\Supp$ (also called the \textit{multi-cut regime} in the case \LogU). The precise assumptions are listed in Section \ref{sec:assumptions}. 

In the Coulomb case, the equilibrium measure $\meseq$ can also be interpreted in terms of the solution to a classical {\it obstacle problem} (and in the Riesz case \eqref{rieszgene} with $\d-2\le \s<\d$ a ``fractional obstacle problem"), which is essentially dual to the minimization of $\I_V$, and better studied from the PDE point of view (in particular the regularity of $\meseq $ and of the boundary of its support). For this aspect, see  \cite[Chap. 2]{ln} and references therein.

Frostman's theorem is the basic result of potential theory.
The relations \eqref{EulerLagrange} can be seen as the Euler-Lagrange equations associated to the minimization of $\I_V$. 
 They state that in the static situation, the total  potential, sum of the potential generated by $\meseq$ and the external potential $V$ must be constant in the support of $\meseq$, i.e. in the set where the ``charges" are present.

More generally $\nab( h^{\mu} + V)$ can be seen as the total mean-field force acting on  charges with density $\mu$ (i.e. each particle feels the average  collective force generated by the other particles), and for  the particle to be at rest one needs that force to vanish. Thus  $\nab (h^{\mu}+ V)$ should vanish on the support of $\mu$, in fact the stationarity condition  that formally emerges as the limit for critical points of $\HN$ is 
\begin{equation}
\label{mflstat} \mu \nab (h^\mu+  V) =0.\end{equation}
The problem with this relation is that the product $\mu \nab h^\mu $ does not always make sense, since a priori $\mu$ is only a probability measure and $\nabla h^\mu$ is not necessarily continuous, however, in dimension 2, one can give a weak form of the equation which always makes sense, inspired by Delort's work in fluid mechanics \cite{delort}, cf. \cite[Chap. 13]{livre}.

%Here, the capacity of a set (see \cite{safftotik,ah} or \cite[Sec. 11.15]{liebloss}) is an appropriate notion of size, suffice it to say that $q.e.$ means except on a set of capacity zero, and that a set of null capacity has zero Lebesgue measure (but the converse is not true).

\subsection{Convergence of minimizers}

\begin{theo}We have
\begin{equation}\label{32}
\lim_{N\to \infty} \frac{\min \HN }{N^2}= \min \I_V= \I_V(\meseq)\end{equation}
and if $(x_1, \dots, x_N)$ minimize $\HN$ then 
\begin{equation}\label{cvmes}
\lim_{N\to\infty} \frac{1}{N} \sum_{i=1}^N \delta_{x_i} \rightharpoonup \meseq\end{equation} 
in the weak sense of probability measures. 
\end{theo}This result is usually  attributed to  \cite{choquet}, one may see the proof in  \cite{safftotik} for the logarithmic cases, the general case can be treated exactly in the same way  \cite[Chap. 2]{ln}, and is valid for very general interactions $\g$ (for instance radial decreasing and integrable near $0$). In modern language it can be phrased as a $\Gamma$-convergence result. 
 It can also easily be expressed in terms of convergence of marginals, as a molecular chaos result.

 \subsection{Parallel results for Ginzburg-Landau vortices}

The analogue mean field result and leading order asymptotic expansion of the minimal energy has also been obtained for the two-dimensional Ginzburg-Landau functional of superconductivity \eqref{gl}, see \cite[Chap. 7]{livre}. It is  phrased as the convergence of the vorticity $\nab \times  \langle i\psi, \nab \psi\rangle$, normalized by the proper number of vortices, to   an equilibrium measure, or the solution to an obstacle problem. The analogue of \eqref{mflstat} is also derived for critical points in \cite[Chap. 13]{livre}.
 \subsection{Deterministic dynamics results - problems \textbf{(3)}}For general reference on problems of the form \textbf{(3)} and \textbf{(4)}, we refer to \cite{spohn}. 
 In view of the above discussion, in the dynamical cases \eqref{gf} or \eqref{hf}, one 
 expects as analogue results the convergences of the (time-dependent) empirical measures $\frac{1}{N} \sum_{i =1}^N \delta_{x_i}$ to probability densities $\mu$ that satisfy the limiting mean-field evolutions
 \begin{equation}\label{mfevol}
 \pa_t \mu = \div (\nab (h^{\mu} + V) \mu)\end{equation}
 respectively
 \begin{equation}\label{mfevol2} \pa_t \mu =- \div ( \mathbb{J} \nab( h^{\mu} + V) \mu)
 \end{equation}
 where again $h^{\mu}= \g* \mu$ as in \eqref{defhmu0}.
 These are nonlocal transport equations where the density $\mu$ is transported along the velocity field $-\nab( h^{\mu} +V)$ (respectively $\mathbb{J} \nab (h^{\mu}+V)$) i.e. advected by the mean-field force that the distribution generates.

 In the two-dimensional Coulomb  case \eqref{wlog2d} with $V=0$, \eqref{mfevol2} with $\mathbb{J}$ chosen as the rotation by $\pi/2$  is also well-known as the vorticity form of the incompressible Euler equation, describing the evolution of the vorticity in an ideal fluid, with velocity given by the Biot-Savart law.
As such, this equation is well-studied in this context, and the convergence of solutions of \eqref{hf} to \eqref{mfevol2}, also known as the point-vortex approximation to Euler, has been rigorously proven, see    \cite{scho,ghl}.

As for \eqref{mfevol}, it is a dissipative equation, that can be seen as a  gradient flow on the  space of probability measures equipped with the so-called Wasserstein   $W_2$ (or Monge-Kantorovitch) metric.  In  the dimension 2 logarithmic case, it was
first introduced by Chapman-Rubinstein-Schatzman \cite{crs} and E \cite{E2} as a formal model for superconductivity, and in that setting the gradient flow description has been made rigorous (see \cite{as}) using the theory of gradient flows in metric spaces of \cite{otto,ags}. The equation can also be studied by PDE methods \cite{lz,sv}, which generalize to the Coulomb interaction in any dimension.  The derivation of this gradient flow equation \eqref{mfevol} from \eqref{glhf} can be guessed  by variational arguments, i.e. ``$\Gamma$-convergence of gradient flows", see \cite{ser4}. 
In the non Coulombic case, i.e. for \eqref{rieszgene}, \eqref{mfevol} is a ``fractional porous medium equation", analyzed in \cite{cv,csv,xz}.

 %The proof  in the third paper relies on  a ``modulated energy" argument  which consists in finding a suitable energy, modelled on the Ginzburg-Landau energy, which measures the distance to the desired limiting solution, and for which a Gronwall inequality can be shown to hold. 
%With a similar ``modulated energy" method, the following result holds 
We have the following result which states in slightly informal terms  that the desired convergence holds provided the limiting solution is regular enough.
\begin{theo}[\cite{ser}] \label{thcv}
For any $\d$, any  case \eqref{wlog2d}, \eqref{wlog} or \eqref{rieszgene} with $\d-2 \le s <\d$, let $\{x_i\}$ solve \eqref{gf}, respectively \eqref{hf} with initial data $x_i(0)=x_i^0$. Then if the limit $\mu^0$ of the initial empirical measure $\frac{1}{N}\sum_{i=1}^N \delta_{x_i^0}$  is regular enough so that the  solution $\mu_t$ of \eqref{mfevol}, resp. \eqref{mfevol2}, with initial data $\mu_0$ exists until time $T>0$ and is regular enough, and if the initial condition is well-prepared in the sense that $$\lim_{N\to \infty} \frac{1}{N^2}\HN(\{x_i^0\}) = \iint_{\R^\d\times \R^\d}\g(x-y) d\mu^0(x) d\mu^0(y),$$
then we have that  for all $t\in[0,T)$, 
$$\frac{1}{N}\sum_{i=1}^N \delta_{x_i(t)} \to \mu_t \quad \text{as} \ N\to \infty.$$\end{theo}
Note that the existence of regular enough solutions that exist for all time, provided the initial data is regular enough, is known to hold for all Coulomb cases and all Riesz cases with $\s<\d-1$ \cite{xz}.

The difficulty in proving this convergence result 
 is due to the singularity of the Coulomb interaction combined with the nonlinear character of the product 
$h^{\mu} \mu$ (and its discrete analogue) which prevents from directly taking limits  in the equation. 

Prior results existed for  less singular interactions \cite{hauray,cch,jw2} or in dimension 1 \cite{bo}. Theorem \ref{thcv} was first proven in the dissipative case in dimensions 1 and 2 in \cite{duerinckx}, then in all dimensions and in the conservative case in \cite{ser}.
Both proofs rely on a ``modulated energy" approach inspired from \cite{serf}. It consists in considering a {\it Coulomb-based (or Riesz-based) distance} between probability densities, more precisely the distance defined by 
$$d_{\g}(\mu, \nu)^2= \iint_{\R^\d\times \R^\d} \g(x-y) d(\mu-\nu)(x) d(\mu-\nu)(y)$$
which is a good metric thanks to the particular properties of the Coulomb and Riesz kernels.
One can prove a ``weak-strong" stability result for the limiting equations \eqref{mfevol}, \eqref{mfevol2} in that metric: if $\mu$ is a smooth enough solution to \eqref{mfevol}, resp. \eqref{mfevol2}, and if $\nu$ is any solution to the same equation, then 
\begin{equation} \label{distcoul}d_{\g}(\mu(t), \nu(t))\le  e^{Ct} d_\g(\mu(0), \nu(0)),\end{equation} which is proved by showing a Gronwall inequality.
One may then exploit this stability property by taking $\mu$  to be the smooth enough expected limiting solution, and $\nu$ to be the empirical measure of the solution to the discrete evolution \eqref{gf} or \eqref{hf},  after giving an appropriate {\it renormalized } meaning to the Coulomb distance (which is otherwise infinite) in that setting.  We are able to prove that a relation similar to \eqref{distcoul} holds, thus proving the desired convergence.

The analogue of the rigorous passage from   \eqref{gf} or \eqref{hf} to \eqref{mfevol} or \eqref{mfevol2} was accomplished  at the level of the full parabolic  and Schr\"odinger Ginzburg-Landau PDEs \eqref{glhf} and \eqref{gls} \cite{ks,jsp2,serf}. The method in \cite{serf} relies  as above on a modulated energy argument  which consists in finding
a suitable energy, modelled on the Ginzburg-Landau energy, which measures the distance to
the desired limiting solution, and for which a Gronwall inequality can be shown to hold.

As far as \eqref{newton} is concerned, the limiting equation is formally found to be the Vlasov-Poisson equation 
\be \label{vp}
\pa_t \ro+ v\cdot \nab_x \ro+ \nab(h^{\mu}+V) \cdot \nab_v \ro=0\ee
where $\ro(t,x,v)$ is the density of particles at time $t$ with position $x$ and velocity $v$, 
and $\mu(t,x)= \int \ro(t,x,v) dv$  is the density of particles. 
The rigorous convergence of \eqref{newton} to \eqref{vp}  and propagation of chaos are not proven in all generality (i.e. for all initial data) but it has been established in a statistical sense  (i.e. randomizing the initial condition) and often truncating  the interactions, see  \cite{kiessling2,bp,hj,lazaro,lp,jabinwang} and also the reviews on the topic   \cite{jabin, golse}.

%Overall, much remains open in this class of problems, even at the mean field level and  how to treat singular interactions such as the Coulomb  one is only known in the conservative cases. 

\subsection{Noisy dynamics - problems \textbf {(4)}}
The noise terms in these equations gives rise to an additive Laplacian term in the limiting equations. For instance the limiting equation for \eqref{noise1} is expected to be the McKean equation
\be \label{mkv}
\pa_t \mu= \frac{1}{\beta} \Delta \mu - \div (\nab( h^{\mu}+V) \mu) \ee
and the convergence is known for regular interactions since the seminal work of \cite{mck},
see also the reviews \cite{sznitman,jabin}.

For singular interactions, the situation has been understood for  the one-dimensional logarithmic case \cite{cl}, then for all Riesz  interactions \eqref{rieszgene} \cite{bo2}. Higher dimensions with singular interactions is largely open, but 
recent progress of \cite{jw2} allows to treat  possibly rough but bounded interactions, as well as some Coulomb interactions,   and prove convergence in an appropriate statistical sense.

% It is expected that the noise should help the convergence and propagation of chaos, but an appropriate method still remains elusive.

For the conservative case \eqref{noise2} the limiting equation is 
a viscous conservative equation of the form
\be \label{ns}
\pa_t \mu= \frac{1}{\beta} \Delta \mu- \div (\np (h^{\mu} +V)\mu )\ee which in the two-dimensional logarithmic case \eqref{wlog2d} is the 
 Navier-Stokes equation in vorticity form. The convergence in that particular case was established in \cite{fhm}, while the most general available result is that  of \cite{jw2}.

 For the case of \eqref{noise3}, the limiting equation is the McKean-Vlasov equation 
 \be\label{mkv2}
 \pa_t  \ro+ v\cdot \nab_x \ro+\nab (h^{\mu}+V) \cdot \nab_v \ro- \frac{1}{\beta} \Delta \ro=0\ee
with the same notation as for \eqref{vp}, and convergence  in the case of bounded-gradient kernels is proven in \cite{jabinwang}, see also references therein.

 \subsection{With temperature: statistical mechanics}
 Let us now turn to problem \textbf{(2)} and consider the situation with temperature as described via the Gibbs measure \eqref{gibbs}.
 One can determine that two temperature scaling choices are interesting: the first is taking $\beta$  independent of $N$, the second is taking $\beta_N= \frac{\beta}{N}$ with some fixed $\beta$.
 In the former, which can be considered a ``low temperature" regime, the behavior of the system is still governed by the equilibrium measure $\meseq$. 
 The result can be phrased using the language of Large Deviations Principles (LDP), cf. \cite{dz} for definitions and reference.

\begin{theo}
 \label{LDP}
The sequence $\{\mathbb{P}_{N, \beta}\}_N$ of probability measures on $\mc{P}(\R^\d)$ satisfies a large deviations principle at speed $N^2$ with good rate function $\beta\hat{\I}_V$ where $\hat{\I}_V = \I_V - \min_{\mc{P}(\R^\d)}  \I_V= \I_V - \I_V(\meseq)$. Moreover 
\begin{equation}
\lim_{N\to + \infty} \frac{1}{N^2} \log Z_{N, \beta} = -\beta \I_V(\meseq) = - \beta \min_{\mc{P}(\R^\d)} \I_V.
\end{equation}
\end{theo} 
The concrete meaning of the  LDP is that if $E$ is a subset of the space of probability measures $\mc{P}(\R^\d)$, after identifying configurations $(x_1, \dots, x_N) $ in $(\R^\d)^N$ with their empirical measures $\frac{1}{N} \sum_{i=1}^N \delta_{x_i}$, we may write
\be  \label{heurisldp}
\P_{N, \beta}(E) \approx e^{-\beta N^2 (\min_{E} \I_V - \min \I_V)}, \ee 
which in view of the uniqueness of the minimizer of $\I_V$ implies that configurations whose empirical measure does not converge to $\meseq$ as $N\to \infty$ have exponentially decaying probability.  In other words the Gibbs measure concentrates as $N\to\infty$ on configurations for which the empirical measure is very close to $\meseq$, i.e. the temperature has no effect on the mean-field behavior.

This result was proven in the logarithmic cases in \cite{hiaipetz} (in dimension 2),  \cite{bg} (in dimension $1$) and  \cite{bz} (in dimension 2) for the particular case of a quadratic potential (and $\beta = 2$), see also \cite{berman2} with results for general powers of the determinant 
 in the setting of multidimensional complex manifolds, or  \cite{cgz} which recently treated more general singular $\g$'s and $V$'s.    This result is actually valid  in any dimension, and is not at all specific to the Coulomb interaction (the proof works as well for more general interaction potentials, see \cite{ln}). 

In the high-temperature regime $\beta_N=\frac{\beta}{N}$, the temperature is felt at leading order and  brings an entropy term. 
 More precisely there is a temperature-dependent equilibrium measure $\mu_{V,\beta}$ which is the unique minimizer of 
 \begin{equation}
 I_{V,\beta}(\mu)= \beta \I_V(\mu) + \int \mu \log \mu.\ee
Contrarily to the equilibrium measure, $\mu_{V,\beta}$ is not compactly supported, but decays exponentially fast at infinity.
This mean-field behavior and convergence of marginals was first established for logarithmic interactions \cite{kiessling,CLMP} (see \cite{messerspohn} for the case of regular interactions) using an approach based on de Finetti's theorem. 
In the language of Large Deviations, the same LDP as above then holds with rate function  $I_{V,\beta}-\min I_{V,\beta}$, and   the Gibbs measure now concentrates as $N\to \infty$ on a neighborhood of $\mu_{V,\beta}$, for a proof see \cite{garcia}. Again the Coulomb nature of the interaction is not really needed.
One can also refer to \cite{nicolas,nicolas2} for the mean-field  and chaos aspects with a particular focus on their adaptation to the  quantum setting.

\section{Beyond the mean field limit : next order study}\label{sec3}

We have seen that studying systems with Coulomb (or more general) interactions at  leading order leads to a good  understanding of their limiting macroscopic behavior. One would like to go further and describe their microscopic behavior, at the scale of the typical inter-distance between the points, $N^{-1/\d}$.  This in fact comes as a by-product of a next-to-leading order description of the energy $\HN$, which also comes together with a next-to-leading order expansion of the free energy in the case \eqref{gibbs}.

Thinking of energy minimizers or of typical configurations under \eqref{gibbs}, 
since one already knows that $\sum_{i=1}^N\delta_{x_i}- N\meseq$ is small, one knows that 
the so-called {\it discrepancy} in balls $B_r(x)$ for instance, defined as   $$D(x,r):= \int_{B_r(x) } \sum_{i=1}^N \delta_{x_i} - N  \, d\meseq$$ is  $o(r^\d N)$ as long as $r>0$ is fixed. 
Is this still true at the mesoscopic scales for $r$ of the order $N^{-\alpha}$ with $\alpha<1/\d$? Is it true down to the microscopic scale, i.e. for $r=RN^{-1/\d}$ with $R \gg 1$? Does it hold regardless of the temperature?  This would correspond to a {\it rigidity result}.
Note that point processes with discrepancies growing like the perimeter of the ball have been called {\it hyperuniform} and are of interest to physicists for a variety of applications, cf. \cite{To}, see also \cite{gol} for a review of the link between rigidity and hyperuniformity.  An addition question is: how much of the microscopic behavior  depends on $V$ or in another words is there  a form of universality in this behavior? 
Such questions had only been  answered in details in the one-dimensional case \eqref{wlog} as we will see below.

\subsection{Expanding the energy to next order}
The first step that we will describe is how to expand the energy $\HN$ around the measure $N\meseq$, following the approach 
  initiated in \cite{ss1} and continued in \cite{ss2,rs,ps,lebles}. It relies on a splitting of the energy into a fixed leading order term and a next order term expressed in terms of the charge fluctuations, and on a rewriting of this next order term  via  the ``electric potential" generated by the points.
More precisely, exploiting the quadratic nature of the interaction, and letting $\triangle $ denote the diagonal in $\R^\d\times \R^\d$, let us expand 
 \begin{eqnarray}
\nonumber \HN(x_1,\dots, x_N)  & = &\hal  \sum_{i \neq j} \g(x_i- x_j) + N \sum_{i=1}^N V(x_i)\\
\nonumber & = & \hal \iint_{\triangle^c} \g(x-y) d\Big(\sum_{i=1}^N \delta_{x_i}\Big) (x)  d\Big(\sum_{i=1}^N \delta_{x_i}\Big)(y) + N \int_{\R^\d} V d\Big(\sum_{i=1}^N \delta_{x_i}\Big)(x) \\
\nonumber 
& = &  \frac{N^2}{2} \iint_{\triangle^c} \g(x-y) d\meseq(x) d\meseq(y) + N^2 \int_{\R^\d} V d\meseq
 % \mbox{     } \right\rbrace \mbox{ terms in } \mu_0 
\\ 
\nonumber
 %\left. 
 & + &  N \iint_{\triangle^c} \g(x-y) d\meseq(x) d\Big(\sum_{i=1}^N \delta_{x_i}-N\meseq\Big)(y)
 % \mbox{     } \right\rbrace \mbox{ term in } n\mu_0 \times (\nu_n - n \mu_0) 
+ N \int_{\R^\d} V d \Big(\sum_{i=1}^N \delta_{x_i}-N\meseq\Big) \\ 
%\left. 
\label{finh}
& + & \hal \iint_{\triangle^c} \g(x-y) d\Big(\sum_{i=1}^N \delta_{x_i}-N\meseq\Big)(x) d\Big(\sum_{i=1}^N \delta_{x_i}-N\meseq\Big)(y).
%\mbox{     } \right\rbrace \mbox{ terms in } d(\nu_n - n\mu_0)(x)d(\nu_n - n\mu_0)(y).
\end{eqnarray}
Recalling that $\meseq$ is characterized by \eqref{EulerLagrange}, we see that the middle term 
\begin{multline}
 N \iint \g(x-y) d\meseq(x)d (\sum_{i=1}^N \delta_{x_i}-N\meseq)(y)
 % \mbox{     } \right\rbrace \mbox{ term in } n\mu_0 \times (\nu_n - n \mu_0) 
+ N \int_{\R^\d} V d (\sum_{i=1}^N \delta_{x_i}-N\meseq) \\
= N \int_{\R^\d} (h^{\meseq}+ V)d (\sum_{i=1}^N \delta_{x_i}-N\meseq) \end{multline}
can be considered as vanishing (at least it does if all the points $x_i$ fall in the support  of $\meseq$).
We are then left with 
\be\label{nxt}
\HN(x_1,\dots, x_N)= N^2 \I_V(\meseq)+ F_N^{\meseq}(x_1, \dots, x_N)\ee
with 
\be F_N^{\meseq}(x_1, \dots, x_N)=  \hal\iint_{\triangle^c} \g(x-y) d\Big(\sum_{i=1}^N \delta_{x_i}-N\meseq\Big)(x) d\Big(\sum_{i=1}^N \delta_{x_i}-N\meseq\Big)(y).\ee
The relation \eqref{nxt} is   a next-order expansion of $\HN $ (recall \eqref{32}), valid for arbitrary configurations. The ``next-order energy" $F_N^{\meseq}$  can be seen as the Coulomb energy of the neutral system formed by the  $N$ positive point charges at the $x_i$'s and the  diffuse negative charge $-N\meseq$ of same mass.
To further understand $F_N^{\meseq}$ let us introduce the potential  generated by this system, i.e. 
\be\label{hN}
H_N(x)= \int_{\R^\d}\g(x-y)d \Big(\sum_{i=1}^N \delta_{x_i}-N\meseq\Big)(y)\ee 
(compare with \eqref{defhmu0}) which solves the linear elliptic PDE (in the sense of distributions)
\be\label{eqlapl}
-\Delta H_N= \cd\Big( \sum_{i=1}^N \delta_{x_i}-N\meseq\Big)\ee
and  use for the first time crucially the Coulomb nature of the interaction to write
\begin{multline}\label{fatr}
 \iint_{\triangle^c} \g(x-y) d\Big(\sum_{i=1}^N \delta_{x_i}-N\meseq\Big)(x) d\Big(\sum_{i=1}^N \delta_{x_i}-N\meseq\Big)(y)\\
 \simeq - \frac{1}{\cd} \int_{\R^\d} H_N\Delta H_N= \frac{1}{\cd} \int_{\R^\d} |\nab H_N|^2\end{multline}
 after integrating by parts by Green's formula.
 This computation is in fact incorrect because it ignores the diagonal terms which must be removed from the integral, and yields a divergent integral $\int |\nab H_N|^2$ (it diverges near each point $x_i$ of the configuration).
 However, this computation can be done properly by removing the infinite diagonal terms and ``renormalizing" the infinite integral, replacing $\int |\nab H_N|^2$ by 
 $$\int_{\R^\d} |\nab H_{N,\eta}|^2- N\cd g(\eta)$$ 
 where  we replace $H_N$ by $H_{N,\eta}$, its ``truncation" at level $\eta$ (here $\eta= \alpha N^{-1/\d}$ with $\alpha$ a small fixed number) --- more precisely $H_{N,\eta}$ is  obtained by replacing the Dirac masses in \eqref{hN} by uniform measures of total mass $1$ supported on the sphere $\pa B(x_i, \eta)$ --- and then removing the appropriate divergent part $\cd \g(\eta)$.
 The name \textit{renormalized energy} originates in the work  of Bethuel-Brezis-H\'{e}lein \cite{bbh} in the context of two-dimensional Ginzburg-Landau vortices, where a similar (although different) renormalization procedure was introduced.
 Such a computation allows to replace the double integral, or sum of pairwise interactions of all the charges and ``background", by a single integral, which is local in the potential $H_N$.    This transformation is very useful, and 
 uses crucially the fact that $\g$ is the kernel of a local operator (the Laplacian). 
 
This electric energy $\int_{\R^\d} |\nab H_{N,\eta}|^2$ is coercive
  and can thus serve to control the ``fluctuations" $\sum_{i=1}^N \delta_{x_i} -N\meseq$,
in fact it is formally $\frac{1}{\cd}\| \nab \Delta^{-1}(\sum_{i=1}^N \delta_{x_i}-N\meseq)\|_{L^2}^2$. 
The relations \eqref{nxt}--\eqref{fatr} can be inserted into the Gibbs measure \eqref{gibbs} to yield so-called ``concentration results" in the case with temperature, see \cite{ser3} (for prior such concentration results, see  \cite{mms,BorGui1,chm}).

\subsection{Blow-up and limiting energy}

As we have seen, the configurations we are interested in are concentrated on (or near) the support of $\meseq$ which is a set of macroscopic size and dimension $\d$, and the typical   distance between neighboring points is $N^{-1/\d}$. The next step is then to blow-up the configurations by $N^{1/\d}$ and take the $N\to \infty$ limit in $F_N^{\meseq}$.
This leads us to a renormalized energy that we define just below. It allows to compute  a total Coulomb interaction for an infinite system of discrete point charges in a constant neutralizing background of  fixed density $1$. 
Such a system is often called a {\it jellium} in physics, and is sometimes considered as a toy model for matter, with a uniform electron sea and ions whose positions remain to be optimized. 

From now on, we assume that $\Sigma$, the support of $\meseq$ is a set with a regular boundary and $\meseq(x)$ is a regular density function in $\Sigma$.
Centering at some point $x$ in $\Sigma$, we may blow-up the configuration by setting $x_i'=\(N \meseq(x)\)^{1/\d}(x_i- x)$ for each $i$. This way we expect to have a density of points equal to $1$ after rescaling.
Rescaling and taking $N\to \infty$ in \eqref{eqlapl}, we  are led to  $H_N\to H$ with $H$ solving an equation of the form
  \be -\Delta H= \cd(\C-1)\ee where  $\C$ is a  locally finite  sum of Dirac masses.
  
  \begin{defi} [\cite{ss1,ss2,rs,ps}] The (Coulomb) renormalized energy of $H$  is 
\begin{equation}\label{defW}
\mc{W}(H) := \lim_{\eta\to 0} \mc{W}_\eta(H)
\end{equation}
where we let
\begin{equation} \label{Weta}
\mc{W}_\eta(H) := \limsup_{R \ti} \frac{1}{R^{\d}} \int_{[-\frac{R}{2}, \frac{R}{2}]^\d}  |\nab H_{\eta}|^2 -  \cd \g(\eta)
\end{equation}
and  $H_\eta$ is a truncation of $H$ performed similarly as above.
\\
We define the renormalized energy of a point configuration $\C$ as
\begin{equation}\label{de522}
\W(\mc{C}) := \inf\{\mc{W}(H) \ | \ -\Delta H= \cd(\C-1)\}
\end{equation}
with the convention $\inf (\emptyset) = +\infty$.
\end{defi}
It is not a priori clear how to define a  total Coulomb interaction of such a  jellium system, because of the infinite size of the system and  because of its lack of local charge neutrality.
The definitions we presented avoid  having to go through computing  the sum of pairwise interactions between particles (it would not even  be clear how to sum them), but instead replace it with (renormalized variants of) the extensive quantity $\int |\nab H|^2$.

The energy $\mathbb W$ can be proven to be bounded below and to have a minimizer;  moreover, its minimum can be achieved as the limit of energies of 
 periodic configurations (with larger and larger period), for all these aspects see for instance \cite{ln}.
 
 \subsection{Cristallization questions for minimizers}
 Determining the value of  $\min \W$  is an open question, with the exception of the  one-dimensional analogues for which the minimum is achieved at the lattice $\mathbb{Z}$  \cite{ss2,leble}.
 
   The only question that we can completely answer so far is that of the minimization over the restricted class of  lattice configurations in dimension $\d=2$, i.e. configurations which are 
 exactly a lattice $\mathbb{Z} \vec{u} + \mathbb{Z} \vec{v}$
 with $det(\vec{u}, \vec{v}) = 1$.
\begin{theo}\label{minimisationreseau} \mbox{}
The minimum of  $\W$ over lattices of volume 1 in dimension 2 is achieved uniquely by the   triangular   lattice.
\end{theo}
Here the triangular lattice means $ \mathbb{Z} + \mathbb{Z} e^{i \pi/3}$, properly scaled, i.e. what is called the Abrikosov lattice in the context of superconductivity. This result is essentially equivalent (see \cite{osp,chiu}) to a result on the minimization of the Epstein $\zeta$ function of the lattice
$$\zeta_s(\Lambda):= \sum_{p\in \Lambda\backslash \{0\}} \frac{1}{|p|^s}$$
 proven in the 50's by Cassels, Rankin, Ennola, Diananda, cf. \cite{montgomery} and references therein. It corresponds to the minimization of the ``height" of flat tori, in the sense of Arakelov geometry.  
 In dimension $\d\ge 3$ the minimization of $\W$ restricted to the class of lattices is   an open question, except in dimensions 4, 8 and 24 where a strict local minimizer is known  \cite{sarns} (it is the $E_8$ lattice in dimension 8 and  the Leech lattice in dimension 24, which were already mentioned before).

One may ask  whether the triangular lattice does achieve the global minimum of $\W$ in dimension 2.  The fact that the Abrikosov lattice is observed in superconductors, combined with the fact  that $\W$ can be derived  as 
 the limiting minimization problem of Ginzburg-Landau \cite{ssgl}, justify  conjecturing this.
  \begin{conjecture}\label{conj}
The triangular lattice is a global minimizer of $\W$ in dimension 2.
\end{conjecture}

It was also recently proven in \cite{betermin} that this conjecture is equivalent to a conjecture of Brauchart-Hardin-Saff \cite{bhs} on the next order term in the asymptotic expansion of the minimal logarithmic energy on the sphere (an important problem in approximation theory, also related to Smale's ``7th problem for the 21st century"), which is obtained by  formal analytic continuation, hence by very different arguments. 
In addition, the result of \cite{cs} essentially yields the local minimality of the triangular lattice within all periodic (with possibly large period) configurations.

Note that the triangular lattice, the $E_8$ lattice in dimension 8 and Leech lattice in dimension 24, mentioned above, are also  conjectured by Cohn-Kumar \cite{ck} to have  universally minimizing properties i.e. to be the minimizer for a broad class of interactions. 
The proof of this conjecture in dimensions 8 and 24 was recently announced by the authors of \cite{ckrmv}, and it should imply that these lattices also minimize $\W$.

 One may expect that in general low dimensions, the minimum of $\W$ is achieved by some particular lattice. Folklore knowledge is that  lattices are not minimizing in large enough dimensions, as indicated by the situation for the  sphere packing problem mentioned above.

These questions  belongs to the more general family of crystallization problems, see \cite{blanclewin} for a review. A typical such question is,  given an interaction kernel $\g$ in any dimension, to determine the point positions that minimize
$$\sum_{i\neq j} \g(x_i-x_j)$$ (with some kind of boundary condition), or rather
$$\lim_{R\to \infty}\frac{1}{|B_R|}\sum_{i\neq j, x_i, x_j \in B_R} \g(x_i-x_j),$$
and to determine whether the minimizing configurations are lattices.  Such questions are fundamental in order to understand
the  cristalline structure of matter. 
There are very few positive results in that direction in the literature, with the exception of  \cite{theil} generalizing \cite{radin} for a class of very  short range Lennard-Jones potentials, which is why the resolution of the sphere packing problem and the Cohn-Kumar conjecture are such breakthroughs.

\subsection{Convergence results for minimizers}

Given a (sequence of) configuration(s) $(x_1, \dots, x_N)$, we examine as mentioned before the blow-up point configurations $\{ (\muv(x)N)^{1/d}(x_i-x)\}$ and their infinite limits $\mathcal{C}$. We also need to let the blow-up center $x$ vary over $\Sigma$, the support of $\meseq$. 
Averaging near the blow-up center $x$ yields a ``point process" $P_N^x$: a point process is precisely 
defined as a 
 probability   distribution on the space of  possibly infinite point configurations, denoted $\config$. 
 Here the point process $P_N^x$ is essentially  the Dirac mass at the blown-up configuration $\{ (\muv(x)N)^{1/d}(x_i-x)\}$. This way, we form a   ``tagged point process" $P_N$ (where the tag is the memory of the blow-up center), probability on $\Sigma \times \config$, whose ``slices" are the $P_N^x$.
 Taking limits $N\to \infty $ (up to subsequences), we obtain limiting tagged point processes $P$, which are   all  stationary, i.e. translation-invariant.  We may also define the renormalized Coulomb energy at the level of tagged point processes as  
$$\bttW(P) :=  \frac{1}{2\cd}\int_{\Sigma}\int  \mathbb W (\mathcal{C}) d P^{x}(\mathcal{C})  dx.$$
 
In view of \eqref{nxt} and the previous discussion, we may expect the following informally stated result (which we state only in the Coulomb cases, for extensions to \eqref{wlog} see \cite{ss2} and to \eqref{rieszgene} see \cite{ps}).
 \begin{theo}[\cite{ss1,rs}] \label{theoW} Consider configurations such that $$\HN(x_1, \dots, x_N) - N^2 \I_V(\muv)\le C N^{2-\frac{2}{\d}}.$$ Then up to extraction $P_N$ converges to some $P$ and 
 \be \label{expH} \HN(x_1,\dots,x_N)\simeq N^2 \I_V(\meseq)+  N^{2-\frac{2}{\d}}  \bttW(P) +o(  N^{2-\frac{2}{\d}})
\ee\footnote{In dimension $\d=2$, there is an additional additive term $\frac{N}{4} \log N$ in both relations}
and in particular 
\be \label{formin} \min \HN= N^2 \I_V(\meseq)+   N^{2-\frac{2}{\d}} \min \bttW+o(  N^{2-\frac{2}{\d}}).\ee \end{theo}
 Since $\bttW $ is an average of $\W$, the result \eqref{formin} can be read as: after suitable blow-up around a point $x$, for a.e. $x\in \Sigma$, the minimizing configurations converge to minimizers of $\W$. If one believes minimizers of $\W$ to ressemble lattices, then it means that minimizers of $\HN$ should do so as well. In any case, $\W$ can distinguish between different lattices  (in  dimension $2$, the triangular lattice has less energy than the square lattice) and we expect $\W$ to be a good quantitative measure of disorder of a configuration (see \cite{bs}).

 The analogous result was proven in \cite{ssgl} for the vortices in minimizers of the Ginzburg-Landau energy \eqref{gl}: they also converge after blow-up to minimizers of $\W$, providing a first rigorous justification of the Abrikosov lattice observed in experiments, modulo Conjecture \ref{conj}. The same result was also obtained in \cite{gms2} for a two-dimensional model of small charged droplets interacting logarithmically called the Ohta-Kawasaki model -- a sort of variant of Gamov's liquid drop model, after the corresponding mean-field limit results was established in \cite{gms1}.

One advantage of the above theorem is that it is valid for generic configurations and not just for minimizers. When using the minimality, better ``rigidity results" (as alluded to above) of minimizers can be proven: points are separated by $\frac{C}{(N\|\muv\|_{\infty} )^{1/d}}$ for some fixed $C>0$  and there is uniform distribution of points and energy, down to the microscopic scale, see \cite{ps,rns,PRN}.

Theorem \ref{theoW} relies on two ingredients which serve to prove respectively a lower bound and an upper bound for the next-order energy. The first is a general method for proving lower bounds for energies which have two instrinsic scales (here the macroscopic scale $1$ and the microscopic scale $N^{-1/\d}$) and which is handled via the introduction of the probability measures on point patterns $P_N$ described above. This method (see \cite{ss1,ln}), inspired by Varadhan, is reminiscent of Young measures and of \cite{am}. 
%It can be applied more generally, for instance to random homogenization \cite{bss}.
The second is a ``screening procedure" which allows to exploit the local nature of the next-order energy  expressed in terms of $H_N$, to paste together configurations given over large microscopic cubes and compute their next-order energy additively. To do so, we need to modify the configuration in  a neighborhood  of the boundary of the cube so as to make the cube neutral in charge and to make 
$\nab H_N$ tangent to the boundary.  This effectively screens the configuration in each cube in the sense that it makes the interaction between the different cubes vanish, so that the energy $\int |\nab H_N|^2$ becomes proportional to the volume. One needs to show that this modification can be made while altering only a negligible fraction of the points and a negligible amount of the energy. This construction is reminiscent of \cite{aco}. It is here crucial that the interaction is Coulomb so that the energy is expressed by a local function of $H_N$, which itself solves an elliptic PDE, making it possible to use the toolbox on estimates for such \medskip PDEs.

The next order study has not at all been touched in the case of dynamics, but it has been tackled in the statistical mechanics setting of \eqref{gibbs}. 
\subsection{Next-order with temperature}
Here the interesting temperature regime (to see nontrivial temperature effects) turns out to be  $\beta_N=\beta N^{\frac{2}{\d}-1}$.

In contrast to the macroscopic result, several observations (e.g. by numerical simulation, see Figure \ref{fig11}) suggest that the behavior of the system at the microscopic scale  depends heavily on $\beta$, and one would like to describe this more precisely.
\begin{figure}[h!] \label{fig:fig}
\begin{minipage}[c]{.46\linewidth}
\begin{center}
\includegraphics[scale=0.18]{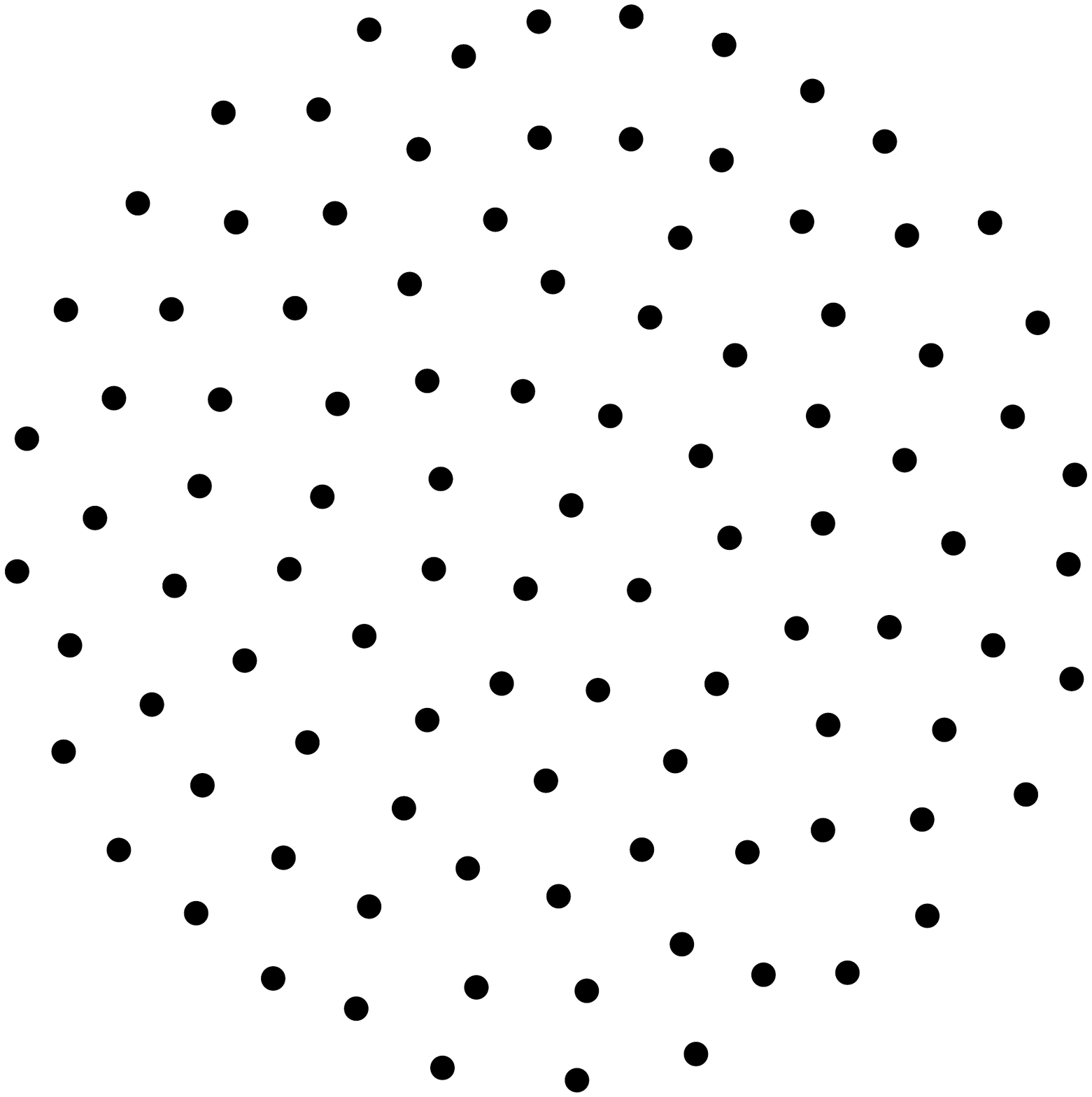} 
\end{center}
\end{minipage}
\begin{minipage}[c]{.46\linewidth}
\begin{center}
\includegraphics[scale=0.19]{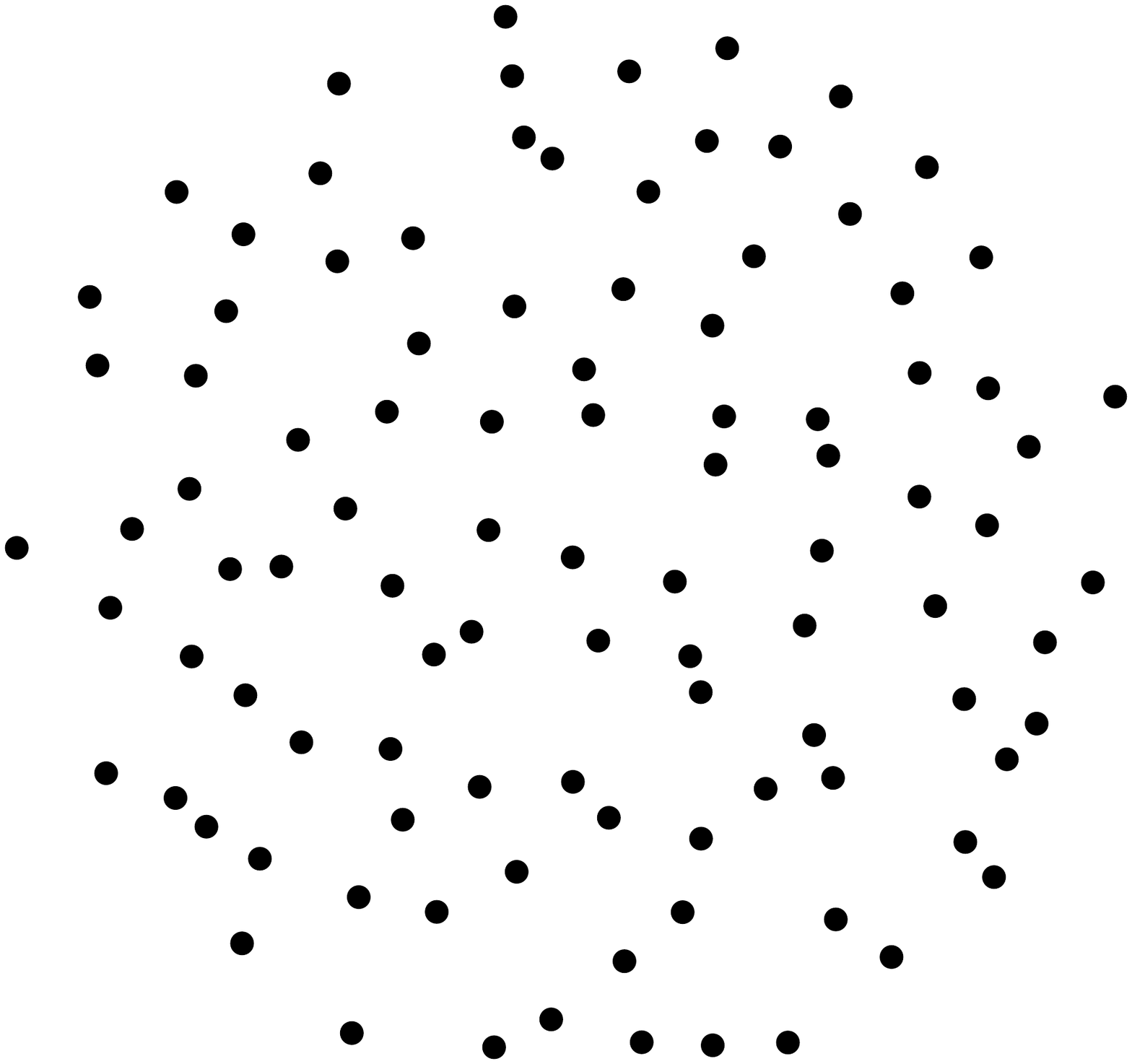} 
\end{center}
\end{minipage}
\vspace{-0.5cm}
\caption{Case \LogD \ with $N = 100$ and $V(x) = |x|^2$, for $\beta = 400$ (left) and $\beta = 5$ (right).}\label{fig11}
\end{figure}
In the particular case of \eqref{wlog} or \eqref{wlog2d} with $\beta=2$, which both arise in Random Matrix Theory, many things can be computed explicitly, and expansions of $\log \ZNbeta$ as $N\to \infty$, Central Limit Theorems for linear statistics, universality in $V$  (after suitable rescaling) of the microscopic behavior and local statistics of the points, are known  \cite{joha,shch,BorGui1,BorGui2,bey1,bey2,bfg,bl}. Generalizing such results to higher dimensions and all $\beta$'s is a significant challenge.

%In all the cases, one wants to understand precisely how the behavior depends on $\beta$, but also on $V$. It is believed that at the macroscopic and microscopic levels, in the logarithmic cases  the behavior is independent on $V$, a  universality feature.

\subsection{Large Deviations Principle}A first approach consists in following the path taken for minimizers and using the next-order expansion of $\HN$ given in \eqref{expH}. This expansion can be formally inserted into \eqref{gibbs}, however this is not sufficient: to get a complete result, one needs to understand precisely how much volume in configuration space $(\R^\d)^N$ is occupied near a given tagged point process $P$ --- this will give rise to an entropy term ---  and how much error (in both volume and energy) the screening construction creates.  At the end we obtain a Large Deviations Principle expressed at the level of the microscopic point processes $P$, instead of the macroscopic empirical measures $\mu$ in Theorem \ref{LDP}.  This  is sometimes called ``type-III large deviations" or large deviations at the level of empirical fields. 
Such results  can be found in \cite{varadhansf}, \cite{follmersf}, the relative specific entropy that we will use  is formalized in \cite{FollmerOrey} (for the non-interacting discrete case), \cite{Georgii1} (for the interacting discrete case) and \cite{georgiz} (for the interacting continuous case).

 To state the result precisely, we need to introduce the Poisson point process with intensity~$1$, denoted $\Poisson$, as the point process characterized by the fact that for any bounded Borel set $B$ in $\R^\d$
$$\Pi \( N(B)= n\)= \frac{|B|^n}{n!}e^{-|B|}$$
where $N(B)$ denotes the number of points in $B$. The expectation of the number of points in $B$ can then be computed to be $|B|$, and one also observes that the number of points in two disjoint sets are independent, thus the points ``don't interact", see Figure \ref{fig2} for a picture.  The ``specific" relative entropy $\bERS$ with respect to $\Poisson$ refers to the fact that it has to be computed taking an infinite volume limit, see \cite{seppalainen} for a precise definition. One can just think that it measures how close the point process is to the Poisson one.

\begin{center}
\begin{figure}
\includegraphics[height=5cm]{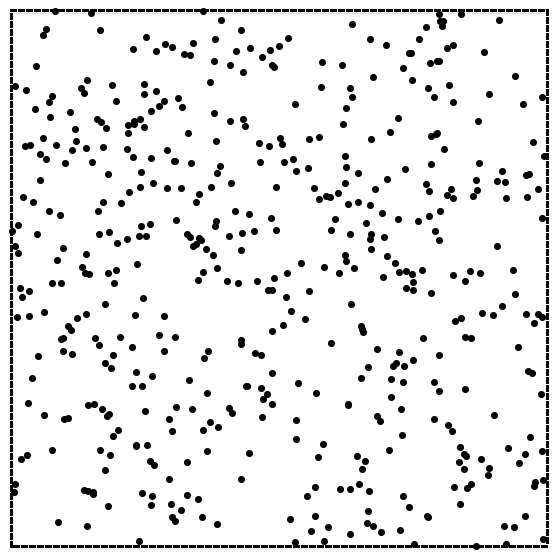}
\includegraphics[height=5cm]{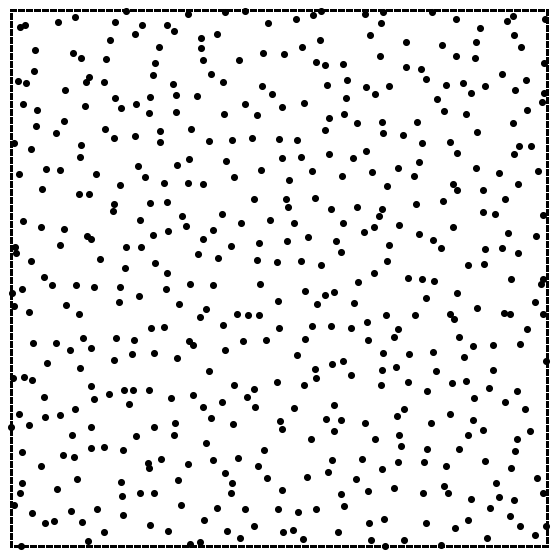}
\caption{Simulation of the Poisson point process with intensity 1 (left), and the Ginibre point process with intensity 1 (right)}
\label{fig2}
\end{figure}
\end{center}

For any $\beta > 0$, we then define a free energy functional $\fbarbeta$ as
\begin{equation}
\label{def:bfbeta} \fbarbeta(P) := \frac{\beta}{2} \bttW(P) + \bERS[P|\Poisson].
\end{equation} 

\begin{theo}[\cite{lebles}] \label{TheoLDP} Under suitable assumptions,  for any $\beta>0$ a Large Deviations Principle at speed $N$ with good rate function $\fbarbeta - \inf \fbarbeta$ holds in the sense that 
$$\PNbeta(P_N\simeq P)\simeq e^{-N(\fbarbeta(P)-\inf \fbarbeta)}$$
\end{theo}

This way, the  the Gibbs measure $\PNbeta$ concentrates on microscopic point processes which minimize  $  \fbarbeta$.  
 This minimization problem corresponds to some balancing (depending on $\beta$) between  $\bttW$, which prefers order of the configurations (and expectedly crystallization in low dimensions), and the relative entropy term which measures the distance to the Poisson process, thus prefers microscopic disorder and decorrelation between the points. As $\beta \to 0$, or temperature gets very large, the entropy term dominates and one can prove \cite{leble} that the minimizer of $\fbarbeta$ converges to the Poisson process. On the contrary, when $\beta \to \infty$, the $\mathbb{W}$ term dominates, and prefers regular and rigid configurations. (In the case \eqref{wlog}  where the minimum of $\mathbb{W}$ is known to be achieved by the lattice, this can be made into a complete proof of crystallization as $\beta \to \infty$, cf. \cite{leble}). 
When $\beta $ is intermediate then both terms are important and one does not expect crystallization in that sense nor complete decorrelation. 
For separation results analogous to those quoted about minimizers, one may see \cite{ameur} and references therein.

The existence of a minimizer to $\fbarbeta$ is  known, it is certainly nonunique due to the rotational invariance of the problem, but it is not known whether it is unique modulo rotations, nor is the existence of a limiting point process $P$ (independent of the subsequence) in general.  The latter is  however known to exist in certain ensembles arising in random matrix theory: for \eqref{wlog} for any $\beta$, it is the so-called sine-$\beta$ process  \cite{ks,vv}, and for \eqref{wlog2d} for $\beta=2$ and $V$ quadratic, it is the Ginibre point process  \cite{ginibre}, shown in Figure \ref{fig2}.  It was also shown to exist for the jellium for small $\beta$ in \cite{imbrie}.
A consequence of Theorem \ref{TheoLDP} is to provide a variational interpretation to these point processes. One may hope to understand phase-transitions at the level of these processes, possibly via this variational interpretation, however this is completely open.
While in dimension $1$, the point process is expected to always be unique, in dimension $2$, phase-transitions and symmetry breaking in positional or orientational order may happen. One would also like to understand  the decay of the two-point correlation function and its possible change in rate, corresponding to a phase-transition. In the one-dimensional logarithmic case, the limits of the correlation functions are computed for rational $\beta$'s  \cite{forres2} and indicate a  phase-transition.
\\

 A  second corollary  obtained as a by-product of Theorem \ref{TheoLDP} 
 is the  
  existence of a next order 
  expansion of the free energy $-\beta^{-1}\log \ZNbeta$.

\begin{coro}[\cite{lebles}]\label{corothermo}
 \begin{equation}\label{logzriesz}
-\beta^{-1}  \log \ZNbeta= N^{1+\frac{2}{\d} } \I_V(\muv) +N \min \mathcal \fbarbeta+o(N)
 \end{equation} in the cases \eqref{kernel}; and in the cases \eqref{wlog2d},  \eqref{wlog},
 $$-\beta^{-1} \log \ZNbeta=  N^2 \mathcal \I_V (\muv) -\frac{ N}{2\d}\log N+N \min \fbarbeta +o(N)$$  or more explicitly
 \begin{multline}\label{logz}
-\beta^{-1} \log \ZNbeta =  N^2\I_V (\muv) -\frac{ N}{2\d}\log N+ N C_\beta +N \left( \frac{1}{\beta}-\frac{1}{ 2\d}\right) \int_\Sigma \muv(x)\log \muv(x) \, dx + o(N),
 \end{multline} where $C_\beta$ depends only on $\beta$, but not on $V$.
 \end{coro}
  This  formulae are to be compared with the  results  of \cite{shch,BorGui1,BorGui2,bfg} in the \LogU\ case, the semi-rigorous formulae in \cite{zv} in the dimension 2 Coulomb case, and are the best-known information on the free energy otherwise. We recall that understanding the free energy is fundamental  for the description of the properties of the system. 
 For instance, the  explicit dependence in $V$ exhibited in \eqref{logz}  will be the key to proving the result of the next section.

Finally, note that  a similar result to the above theorem and corollary can be obtained in the case of the two-dimensional two-component plasma alluded to in Section \ref{2comp}, see \cite{2D2CP}.

\subsection{A Central Limit Theorem for fluctuations}  
Another approach to understanding the rigidity of configurations and how it depends on the temperature is to examine the behavior of the linear statistics of the fluctuations, i.e. consider, for a regular test function $f$, the quantity 
$$\sum_{i=1}^N f(x_i)- N \int fd \meseq.$$

\begin{theo}[\cite{ls2}]  \label{theoclt}
In the case \eqref{wlog2d}, assume $V\in C^{4}$ and the previous assumptions on $\mu_V$ and $\partial \Sigma$, and let $f\in C_c^{4}(\R^2)$ or $C_c^{3}(\Sigma)$.  
If $\Sigma$ has $m\ge 2$ connected components $\Sigma_i$, add $m-1$ conditions
$\int_{\pa \Sigma_i } \Delta f^\Sigma=0$ 
where $f^\Sigma$ is the harmonic extension of $f$ outside $\Sigma$.
Then 
$$\sum_{i=1}^N f(x_i)- N \int_{\Sigma }f \, d\muv$$
converges  in law as $N \to \infty$ to  a Gaussian distribution with 
$$\text{mean = }  \frac{1}{2\pi}\( \frac{1}{\beta}-\frac{1}{4}\) \int_{\R^2} \Delta f \, (\indic_{\Sigma} + \log \Delta V )^\Sigma  \qquad \text{variance= }\frac{1}{2\pi \beta} \int_{\R^2} |\nab f^\Sigma|^2.$$  \\
 The result can  moreover be localized with $f$ supported on any mesoscale $N^{-\alpha}$, $\alpha <\frac{1}{2},$ and it  is true as well for energy minimizers, taking formally $\beta=\infty$. 
\end{theo}
This result can be interpreted in terms of the convergence of $H_N$ (of \eqref{hN})   to a suitable so-called  ``Gaussian Free Field", a sort of two-dimensional analogue of  Brownian motion. 
This theorem shows that if $f$ is smooth enough, the fluctuations of linear statistics are  typically of order $1$, i.e. much smaller than the sum of $N$ iid random variables which is typically or order $\sqrt{N}$. This a manifestation of rigidity, which even holds down to the mesoscales.
Note that the regularity of $f$ is necessary, the result is false if $f$ is discontinuous, however the precise threshhold of regularity is not known.

 In dimension 1, this theorem was first proven in \cite{joha} for polynomial $V$ and $f$ analytic. It was later generalized in \cite{shch,BorGui1,BorGui2,bl,llw,bls}.
  In dimension 2, 
 this  result was proven for the determinantal case $\beta=2$, first  in \cite{ridervirag} (for $V$ quadratic), 
 \cite{berman3} assuming just $f\in C^1$, 
 and then \cite{ahm} under analyticity assumptions. It was then proven for all $\beta$ simultaneously as \cite{ls2} in  \cite{bbny2}, with $f$  assumed to be supported in $\Sigma$. 
 The approach for proving such results has generally been based on Dyson-Schwinger (or ``loop") equations. 
    
      If the extra conditions do not hold, then the CLT is not expected to hold. Rather, the limit should be a Gaussian convolved with a discrete Gaussian variable, as shown in the \LogU \ case in \cite{BorGui2}.
 \\
  
To prove Theorem \ref{theoclt}, following the approach pioneered by Johansson \cite{joha}, we compute the Laplace transform of these linear statistics and
 see that it reduces to understanding the ratio of two partition functions, the original one and that of a Coulomb gas with 
 potential $V $ replaced by $V_t=V+t f$ with $t$ small. Thanks to \cite{serser} the variation of the equilibrium measure  associated to this replacement is well understood. We are then able to 
 leverage on the expansion of the partition function  of \eqref{logz} to compute the desired ratio, using also a change of variables which is a transport map between the equilibrium measure $\mueq$ and  the perturbed equilibrium measure.
   Note that the use of changes of variables in this context is not new, cf. \cite{joha,BorGui1,shch,bfg}.  In our approach, it essentially replaces the use of the loop or Dyson-Schwinger equations.

\subsection{More general interactions}
It  remains to understand how much of the  behavior we described  are really specific to Coulomb interactions. Already Theorems \ref{theoW} and \ref{TheoLDP} were shown in \cite{ps,lebles} to hold for the more general Riesz interactions with $\d-2\le s <\d$. This is thanks to the fact that the Riesz kernel is the kernel for a fractional Laplacian, which is not a local operator but can be interpreted as a local operator after adding one spatial dimension, according to the procedure of Caffarelli-Silvestre \cite{caffsilvestre}. The results of Theorem~\ref{TheoLDP}  are also valid in  the {\it hypersingular} Riesz interactions $\s >\d$ (see \cite{hlss}), where the kernel is very singular but also  decays very fast. 
The Gaussian behavior of the fluctuations seen  in   Theorem~\ref{theoclt} is for now proved only in the logarithmic cases, but it remains to show whether it holds for more general Coulomb cases and even possibly more general interactions as well.
\vskip 1cm

{\bf Ackowledgements:} I am very grateful to Mitia Duerinckx, Thomas Lebl\'e, Mathieu Lewin and Nicolas Rougerie for their helpful comments and suggestions on the first version of this text. I also thank Catherine Goldstein for  the historical references and Alon Nishry for the pictures of Figure \ref{fig2}.

\end{document}